\documentclass[12pt]{article}
\newcommand{\BE}{\begin{eqnarray}}
\newcommand{\EE}{\end{eqnarray}}
\usepackage{amssymb}
\usepackage{amsmath}
\usepackage{array}
\usepackage{latexsym}

\begin{document}
\vspace*{-.6in}
\thispagestyle{empty}
\begin{flushright}
CALT-68-2423
\end{flushright}
\baselineskip = 18pt

\vspace{1.5in} {\Large
\begin{center}
Two-Graviton Interaction in PP-Wave Background in Matrix
Theory and Supergravity
\end{center}} \vspace{.5in}

\begin{center}
Hok Kong Lee and Xinkai Wu
\\
\emph{California Institute of Technology\\ Pasadena, CA  91125, USA\\ hok@theory.caltech.edu, xinkaiwu@theory.caltech.edu}
\end{center}
\vspace{1in}

\begin{center}
\textbf{Abstract}
\end{center}
\begin{quotation}
\noindent We compute the two-body one-loop effective action for the matrix
theory in the pp-wave background, and compare it to the effective
action on the supergravity side in the same background. Agreement is
found for the effective actions on both sides. This points to the
existence of a supersymmetric nonrenormalization theorem in the
pp-wave background. 
\end{quotation}

\newpage

\pagenumbering{arabic}

\section{Introduction}

The Matrix model of M-theory proposed by ~\cite{BFSS, Susskind} is
believed to give a  non-perturbative description of quantum
gravity. In the original paper ~\cite{BFSS},  the authors computed
graviton scattering in flat space using the Matrix model and  found
exact agreement with 11-dimensional Supergravity, which is believed to
be the  low energy limit of M-theory. The result also suggested the
existence of a  nonrenormalization theorem for the $v^4$ term in the
effective action, which was  subsequently confirmed by the study of
supersymmetry in ~\cite{PSS1, PSS2, PSS3, HKS,  LOWE, NP, KM}. Since
then, more detailed investigations have been performed in  flat space
~\cite{Becker, BB1, BB2, BB3, BBPT, OY1, OY2, Okawa}.  The Matrix
model in a weakly curved background was proposed by Taylor and Van
Raamsdonk in  ~\cite{TRCB}. The case of a space weakly curved in the
transverse directions was checked explicitly in  ~\cite{TR}.

The Matrix Theory action in the pp-wave background was proposed in
~\cite{BMN}. Since this  new action is exact in this curved
background, it could provide further tests of the  Matrix conjecture
beyond the proposal on weakly curved backgrounds in ~\cite{TRCB}. In
addtion  it is different from the case in ~\cite{TR}, because the
background  metric now has a nontrivial $g_{++}$ component.

The goal of this paper is to compare two-graviton scattering in the
pp-wave background  in Matrix Theory and Supergravity. An agreement
will provide evidence for: (1) the existence of a supersymmetric
nonrenormalization theorem in pp-wave background for the terms
compared; (2) Matrix model as a description of M-theory in the pp-wave
background. As pointed out by \cite{DSO, DO}, Matrix Theory in a
generic curved background is not expected to agree with
Supergravity. In the pp-wave case, however, we do find precise
agreement as will be shown in this paper. This is likely to be a
result of the large number of supersymmetries of the pp-wave
background.

The paper is organized as follows. In Section \ref{Rev}, we briefly
review the  previously known results in flat space and in weakly
curved backgrounds. In Section  \ref{EffP}, the approximations and
limits of our computation are discussed. In  Section~\ref{BFM} through
Section \ref{GenCase}, we present the computation on the  Matrix
Theory side. In Section~\ref{sec:sugra} we present the computation on
the  Supergravity side. In Section~\ref{sec:discussion} we discuss
some possible future  directions. Some of the technical details are
given in the Appendices.

\section{A Brief Review of Known Results}\label{Rev}

In \cite{BB1}, the one-loop effective potential for two gravitons in
flat spacetime  background was computed in Matrix Theory to be:
\begin{eqnarray}
V_{eff}^{1-loop}=\frac{15N_pN_sv^4}{16M^9R^3r^7}\label{flatresult}
\end{eqnarray}
where $N_p$ and $N_s$ are the numbers of D0-branes making up the probe
graviton and source graviton, respectively, $v$ and $r$ are the
transverse relative velocity and distance between them,  $M$ is the
11-dimensional Planck mass, and $R$ is the radius of compactification
in  DLCQ. This effective potential agrees precisely with the
Supergravity result  \cite{BBPT}.

In \cite{TR}, the effective potential for a weakly curved background
with nontrivial  transverse metric components was computed. Again
agreement was found. In fact, the  only modification needed was the
replacement of $r$ by $d$, the geodesic distance  between the two
gravitons.

\section{The Effective Potential, $V_{eff}$}\label{EffP}

In this paper, the main object for comparison on both sides is the
effective potential  $V_{eff}$. The computation is carried out in the
DLCQ formalism, which was proposed in  Susskind's finite N conjecture
\cite{Susskind}, and further elucidated by \cite{Seiberg:1997ad,
Sen:1997we}. In this formalism $x^-$ and $x^- + 2\pi R$  are
identified. $p_-$ is therefore quantized in units of $1/R$.

The implications of such a lightlike compactification, however, are
far from trivial\cite{Pol1}. One such complication arises from the
longitudinal zero modes, which appear to cause perturbative amplitudes
to diverge. In addition, there are concerns that the DLCQ of M-theory
in the low energy limit is not necessarily the DLCQ of 11-dimensional
Supergravity because some exotic degrees of freedom such as membranes
wrapped around the lightlike direction may contribute.

Here we are going to take the viewpoint in \cite{BGL}. Essentially,
the presence of a source exerts a pressure that decompactifies the
region surrounding it, rendering $x^-$ effectively spacelike by
providing a nonzero $g_{--}$ component in the metric.  In the limit of
large N, this bubble of 11-dimensional space expands, and the
approximation of Supergravity as a low energy description is thus
justified. This view is further elucidated in \cite{Pol2}, and we do
not expect new issues to arise in the pp-wave background.

It was also argued in \cite{Pol2} that the perturbative calculations
in Matrix Theory and Supergravity have different regions of validity
($E/M>N^{1/3}$ and $E/M<N^{-1}$). Therefore, in general there is no
reasons why they should match as each effective action is valid only
within its own energy scale.  Thus a mismatch does not immediately
invalidate the Matrix Conjecture. An exact match, however, will point
to the existence of a nonrenormalization theorem, which protects the
terms evaluated from gaining higher loops corrections. If such a
nonrenormalization theorem does exist, then the agreement of both
sides can be viewed as positive evidence for the Matrix Conjecture. It
is with these points in mind that the comparison of the effective
action is made here.

On the Matrix Theory side, the effective potential is computed up to
1-loop. As in flat space, it should correspond to to terms of order
$\kappa_{11}^2$ on the Supergravity side.  The relation
$\kappa_{11}^2=16 \pi^5/M^9$ \cite{BBPT} means only terms of order
$1/M^9$ are relevant on the Matrix Theory side for the purpose of such
comparison.

A natural length scale that arises on the Matrix Theory side is
$1/(M^3 R)^{1/2}$, which for convenience we will denote as
$(\alpha)^{1/2}$.
\footnote{This $\alpha$ should not be confused with the string scale
$\alpha '$.} In addition to the low velocity and large $r$
approximation necessary to facilitate comparison in flat space, we
will assume also:

\begin{eqnarray} \label{Small} 
\frac{\alpha ^2 \mu^2}{r^2}<<1
\end{eqnarray}
where $\mu$ is the $123+$ component of the four-form field strength.

This dimensionless number, as we will see in eqn(\ref{Maction}), is
simply the relative strength of the new terms in the action arising
from the pp-wave background to the quartic terms already present in
flat space. In the opposite limit, $\frac {r^2}{\alpha ^2 \mu^2}<<1 $,
the effective potential on the Matrix Theory side resums to give $1/
\mu$ dependence\footnote{This can be seen in eqn(\ref{SQR}), a typical
term in the effective potential.}, which does not appear possible to
be reproduced on the Supergravity side. In fact, this is nothing
new. A similar issue arises already in flat space, where the effective
potential only matches when we take the small $v$ and large $r$ limit,
or more precisely, by expanding on the small parameter $v \alpha /
r^2$. In other words, even with the existence of a nonrenormalization
theorem, the results on both sides should only be compared at very
large $r$, where Supergravity is applicable.

\section{Background Field Method}\label{BFM}

We will follow the Background Field Method as reviewed
in~\cite{Elvang}. $X$ is expanded into a background field $B$ and a
fluctuating field $Y$, i.e. $X=B+Y$. Only the part of the action that
is  quadratic in $Y$ will be of  interest below.

The Matrix Model action in the DLCQ of M-theory in the maximally
supersymmetric pp-wave background is ~\cite{BMN}:
\begin{eqnarray} \label{Maction}
S=\int dt Tr\Bigg\{\sum_{I=1}^9\frac{1}{2R}(D_0 X^I)^2+i
\psi^TD_0\psi+  \frac{(M^3R)^2}{4R}\sum_{I,J=1}^9[X^I,X^J]^2+(M^3R)
\sum_{J=1}^9  \psi^T\gamma^J[\psi,X^J] \nonumber\\  +
\frac{1}{2R}\left[-(\frac{\mu}{3})^2\sum_{i=1}^3(X^i)^2
-(\frac{\mu}{6})^2\sum_{a=4}^9(X^a)^2\right] -  i
\frac{\mu}{4}\psi^T\gamma_{123}\psi \nonumber\\
-\frac{(M^3R)\mu}{R}i\sum_{i,j,k=1}^3\epsilon_{ijk} Tr(X^iX^jX^k)
\Bigg\}
\end{eqnarray}
where $D_0X=\partial_tX^I-i[X_0,X^I]$

Taking the ratios of any of the $\mu$-dependent terms to the
$\mu$-independent  non-derivative terms gives the parameter in
eqn(\ref{Small}). In other words, the  assumption stated in the
previous section is identical to treating the new terms  arising from
the pp-wave background as a perturbation to flat space. Note that this
is exactly the opposite of the  approximation made in ~\cite{DSR},
where the $\mu$-independent terms are treated  as perturbations to the
$\mu$-dependent terms. While the computation of the  1-loop effective
potential is possible in both limits on the Matrix Theory side, an
agreement with Supergravity is possible only in the large $r$ limit
given in eqn(\ref{Small}).

In what follows, unless stated otherwise, we will always assume the
indices $i$ goes from 1 to 3, $a$ goes from 4 to 9, and $I$ goes from
1 to  9.

In addition to the action above, there are terms arising from the
ghosts and gauge fixing,  which we simply state below:

\begin{equation}
S_{gf}=\int dt Tr \Bigg [ -\frac{1}{2R} (\partial_t X_0 +i [B_i,
X_i])^2  \Bigg]
\end{equation}

\begin{equation}
S_{ghost}=\int dt Tr \Bigg [ \overline{c} \partial_t^2 c - \partial_t
\overline{c}[X_0, c] +  \overline{c}[B^i, [X^i, c]] \Bigg]
\end{equation}
Thus, the complete Matrix Theory action is:

\begin{equation}
S_M=S+S_{gf}+S_{ghost}
\end{equation}

To simplify the notation, we will put $M^3R=1/ \alpha=1$. This factor
can be  restored by dimensonal analysis. It is also convenient to
define  $g^2\equiv R$, which corresponds to a loop counting parameter
in the Matrix  Theory.


\subsection{Expansion About The Backgroud}

The fields $X$, $\psi$, and $c$ are expanded in the following way,
with a  purely bosonic background. Here we set $N_p=N_s=1$, i.e. we deal with $2\times 2$ matrices (much interesting work has been done on matrix quantum mechanics in flat
space, e.g. \cite{Halpern:1997fv, Claudson:1984th} ). We will later restore $N_p$ and $N_s$:
\begin{eqnarray}
X_\mu=B_\mu+g Y_\mu \ &;& \quad \mu=0, 1, 2, ..., 9 \nonumber\\
B_I=\begin{pmatrix} x_I & 0 \\ 0 & 0 \end{pmatrix} \ &;& \quad 
Y_I=\begin{pmatrix}  
\zeta_I & z_I \\ \overline{z}_I & \widetilde{\zeta}_I \end{pmatrix} 
\nonumber \\
B_0=\begin{pmatrix} 0 & 0 \\ 0 & 0 \end{pmatrix} \ &;& \quad 
Y_0=\begin{pmatrix}  
\zeta_0 & z_0 \\ \overline{z}_0 & \widetilde{\zeta}_0 \end{pmatrix} 
\nonumber \\
\psi=\begin{pmatrix} \eta & \theta \\ \overline{\theta} & 
\widetilde{\eta}  
\end{pmatrix} \ &;& \quad
c=\begin{pmatrix} \epsilon & c_1 \\ c_2 & \widetilde{\epsilon} 
\end{pmatrix}   
\nonumber   
\end{eqnarray}      

The above background has the interpretation of one graviton (the
source) sitting at the origin\footnote{Another possible interpretation
is a transverse five brane at the origin \cite{MALD}.}, while another
graviton (the probe) approaches  from the position given by $x^I$ in
the matrix $B$. We will use the shorthand  $r^2=\sum_{I=1}^9 (x^I)^2$.

After a Wick rotation, where we define $S=iS^{(E)}$ and $\tau=it$, and
at the  same time rotating $X_0$ to $iX_0^{(E)}$, the quadratic part
of the action  is: \footnote{For simplicity, all subsequent
superscripts of $(E)$ on the  Euclideanized fluctuation fields will be
omitted.}

\begin{multline} \label{S2b}
{\cal S}_{boson}^{(E)}= \int d\tau \Bigg \{ -\frac{1}{2} \zeta_0
\partial_\tau^2 \zeta_0 -\frac{1}{2} \widetilde{\zeta}_0
\partial_\tau^2 \widetilde{\zeta}_0 + \frac{1}{2} \zeta_i
(-\partial_\tau^2 +(\mu /3)^2) \zeta_i + \frac{1}{2} \zeta_a
(-\partial_\tau^2 +(\mu /6)^2) \zeta_a \\  +\frac{1}{2}
\widetilde{\zeta}_i (-\partial_\tau^2 +(\mu /3)^2) \widetilde{\zeta}_i
+ \widetilde{\zeta}_a (-\partial_\tau^2 +(\mu /6)^2)
\widetilde{\zeta}_a \\  + \overline{z}_0 (-\partial_\tau^2 + r^2) z_0
-2i \partial_\tau x_I  (\overline{z}_I z_0 - \overline{z}_0 z_I) \\  +
\overline{z}_i (-\partial_\tau^2 + r^2 + (\mu /3)^2) z_i +
\overline{z}_a  (-\partial_\tau^2 + r^2 + (\mu /6)^2) z_a - i \mu
\epsilon_{ijk} x_i \overline{z}_j z_k \Bigg \}
\end{multline}

\begin{multline}\label{S2f}
{\cal S}_{fermion}^{(E)}= \int d\tau \Bigg \{ \eta (i \partial_\tau
-i\frac{\mu}{4}  \gamma_{123}) \eta + \widetilde{\eta} (i
\partial_\tau - i \frac{\mu}{4}  \gamma_{123}) \widetilde{\eta} + 2
\overline{\theta}( i \partial_\tau +  x_I  \gamma_I -i
\frac{\mu}{4}\gamma_{123}) \theta \Bigg \}
\end{multline} 

\begin{multline} \label{S2g}
{\cal S}_{ghost}^{(E)}= \int d\tau \Bigg \{ \overline{\epsilon}
\partial_\tau^2 \epsilon +
\overline{\widetilde{\epsilon}}\partial_\tau^2 \widetilde{\epsilon}+
\overline{c}_1 (\partial_\tau^2 - r^2) c_2 + \overline{c_2}
(\partial_\tau^2 -  r^2)c_1 \Bigg \}
\end{multline}

\subsection{The Sum Over Mass}\label{SDM}

The partition function, ${\cal Z}$ of the above action can be computed
as  a product of functional determinants. The 1-loop effective action
$\Gamma$  is then simply related to ${\cal Z}$ via:

\begin{equation}
exp(-\Gamma)={\cal Z}
\end{equation}
The 1-loop effective potential is defined as:

\begin{equation}
\Gamma=-\int d\tau \ V_{eff}
\end{equation}

To first approximation, however, it is not necessary to compute the
functional determinants. As was suggested by Talfjord and Periwal
~\cite{TAF} and Talyor ~\cite{TAY01}, one could deduce the effective
potential by simply evaluating the mass spectrum of the fluctuating
fields. From the masses, the 1-loop contribution to $V_{eff}$ could be
easily deduced using the formula:

\begin{equation}\label{Vformula}
V_{eff}^{1-loop}=-\frac{1}{2}\left(\sum_{real\ bosons}m_b-\sum_{real\
fermions}m_f-\sum_{real\ ghosts} m_g \right)
\end{equation}

The physical reasons for this is that at large distances, i.e. the
limit  where Supergravity is valid, all the string stretching between
the  D0-branes can be assumed to lie in their ground state. This
result can  also be verified using the complete expression for
$V_{eff}$ in terms of  functional determinants. We provide an argument
for this in Appendix A. In what follows, we will omit the superscript
$1-loop$, assuming this is understood. The contribution from tree
level, which does not concern us  here, is simply the Lagrangian with
$X$ replaced by $B$. Both  contributions will be put back together at
the end in eqn(\ref{VGauge2}).

One important point to note is that this method is valid only up to
the  lowest powers of $v$, as is already known in the flat space
case. In flat  space, the above formula reproduces every term
predicted by a  Supergravity computation with the right coefficients,
but the Matrix  Theory corrections to Supergravity, i.e. terms with
even higher powers of  $v$ and $1/r$ which would not be found in
Supergravity, will not come  out with the correct coefficients. In
fact, the parameter $\alpha$ can be treated as  the counting parameter
for this purpose. All terms of order $\alpha^3$, which is  basically
$\kappa_{11}^2$ in the Supergravity language, will be found on the
Supergravity side, but terms on the Matrix Theory side with higher
powers of $\alpha$,  which represent short distance effects, should be
treated as corrections. To  compute them correctly, one needs to make
use of the complete expression in terms of  functional determinants.

For our purpose, however, the above approach is sufficient. We are not
interested in computating the correction to Supergravity, rather we
would  like to check whether the terms already predicted by
Supergravity in the  pp-wave background can be reproduced by a Matrix
Theory calculation.\\

\section{A Simple Case}\label{SimCase}

In the next section we will work out a more efficient method to
compute $V_{eff}$  without  explicitly diagonalizing the mass
matrix. Nevertheless, it is instructive to work out  the simplest case
in a direct approach to get the basic idea of the computation.

In this simple case, we put $x^8=b$ and $x^9=v \tau$, while all the
other $x^I$ are  set to  zero\footnote{Note that by putting all $x^i$
to zero for $i=1,2,3$, we made sure that  in this case  the Myers term
will not contribute to the mass matrix.}. Here $b$ is a constant,
which can be  interpreted as the impact parameter of the approaching
probe graviton towards the  source sitting at  the origin. In this
case, the mass matrix constructed from eqn(\ref{S2b}), (\ref{S2f}) and
(\ref{S2g}) is easily diagonalized to give the mass spectrum listed in
Table \ref{table}.
\begin{table}
\begin{center}
\setlength{\extrarowheight}{4pt}
\begin{tabular}{|m{5cm}|m{5cm}|}
\hline $\mathbf{m^2}$ & \textbf{Fields}\\ \hline $0$ & $\zeta^0$\\
\hline $\mu^2/3^2$ & $\zeta^i\quad;i=1,2,3$\\ \hline $\mu^2/6^2$ &
$\zeta^a\quad;a=4,...,9$\\ \hline $0$ & $\widetilde{\zeta}^0$\\ \hline
$\mu^2/3^2$ & $\widetilde{\zeta}^i\quad;i=1,2,3$\\ \hline $\mu^2/6^2$
& $\widetilde{\zeta}^a\quad;a=4,...,9$\\ \hline $r^2+\mu^2/3^2$ &
$\overline{z}^i, z^i\quad;i=1,2,3$\\ \hline $r^2+\mu^2/6^2$ &
$\overline{z}^a, z^a\quad;a=4,...,8$\\ \hline $r^2+\eta_+$ &
$\overline{z}^0+\overline{z}^9, z^0+z^9$\\ \hline $r^2+\eta_-$ &
$\overline{z}^0-\overline{z}^9, z^0-z^9$\\ \hline $\mu^2/4^2$ & $\eta
\quad (8)$\\ \hline $\mu^2/4^2$ & $\widetilde{\eta} \quad (8)$\\
\hline $r^2+\mu^2/4^2+v$ & $\theta \quad (8)$\\ \hline
$r^2+\mu^2/4^2-v$ & $\overline{\theta} \quad (8)$\\ \hline $0$ &
$\overline{\epsilon}, \epsilon $\\ \hline $0$ &
$\overline{\widetilde{\epsilon}}, \widetilde{\epsilon} $\\ \hline
$r^2$ & $\overline{c}_I, c_I\quad ;I=1,2$\\ \hline
\end{tabular}
\caption{The Mass Spectrum for a Simple Case. The numbers inside the
round brackets indicate the number of physical  degrees of freedom of
the fermions with the given mass. 
$\eta_\pm$ is given by $\frac{1}{2}[\frac{\mu^2}{6^2}\pm\sqrt{(\frac{\mu^2}{6^2})^2+16v^2}]$
.}
\label{table}
\end{center}
\end{table}
It should be noted that the velocity in the table above is measured in
Euclidean time  $\tau$, i.e.  $v=\frac {\partial x}{\partial
\tau}$. In a comparison with Supergravity, a Wick  rotation back into
Minkowski time   $t=-i\tau$ is required, which introduces extra minus
signs in $V_{eff}$.

With the mass spectrum at hand, $V_{eff}$ can be evaluated using
eqn(\ref{Vformula}):

\begin{multline}
V_{eff} =-\frac {1}{2}(2)
(3\frac{\mu}{3}+6\frac{\mu}{6}-8\frac{\mu}{4})-\frac{1}{2}\Bigg\{
6\sqrt{r^2+\mu^2/3^2}+10\sqrt{r^2+\mu^2/6^2}+2\sqrt{r^2+\eta_+} \\ +
2\sqrt{r^2+\eta_-}-8\sqrt{r^2+\mu^2/4^2+v}-8\sqrt{r^2+\mu^2/4^2-v}-4r
\Bigg\}
\end{multline}

At this point it is useful to restore the factors of $M^3R$, which we
denote as $1/ \alpha$. For instance, the first square root term in the
about equation  becomes:

\begin{equation}\label{SQR}
\sqrt{\frac{r^2}{\alpha^2}+\frac{\mu^2}{3^2}}
\end{equation}

This can in turn be written as:

\begin{equation}
\frac{r}{\alpha} \sqrt{1 + \frac{1}{3^2} (\frac{\alpha}{r^2}) (\alpha
\mu^2)}  \nonumber
\end{equation}

The expression for $V_{eff}$ given above, being a Matrix Theory
result, is only  expected to match with Supergravity in the large $r$
limit (if it does at all!). Defining the large $r$ limit by
eqn({\ref{Small}), we can then expand the 1-loop effective potential
in powers of $\alpha^2 \mu^2/r^2$. Thus, expanding $V_{eff}$ gives:

\begin{equation}
V_{eff} =\alpha^3 (\frac{15}{16}\frac{v^4}{r^7}
+\frac{7}{96}\frac{\mu^2v^2}{r^5}+\frac{1}{768} \frac{\mu^4}{r^3}) +
O[\alpha^5]
\end{equation}

Wick rotating $v$, and restoring $N_p, N_s$ gives:\\
\begin{eqnarray}
V_{eff}=\frac{N_p N_s}{M^9 R^3}
(\frac{15}{16}\frac{v^4}{r^7}-\frac{7}{96}\frac{\mu^2v^2}{r^5}+\frac{1}{768}
\frac{\mu^4}{r^3})+O[\alpha^5]
\end{eqnarray}

The $\alpha^3$ terms give the factor  $1/M^9$, which translates into
$\kappa_{11}^2$ in the Supergravity language. This is the order we are
interested in. We throw away the higher powers of $\alpha$ (which are
always accompanied  by powers of $1/r$) because they correspond to
short distance corrections to  Supergravity, just as in flat space.

Here the first term is just the flat space result. The second and the
third term are the interesting ones, with new $\mu^2v^2$ and  $\mu^4$
dependence created by the pp-wave background. A comparison of their
coefficients with Supergravity will show exact agreement.\\

\section{Mass Matrix Computation} \label{MMC}

In the more general cases, when the velocity and the impact parameter
point in arbitrary directions, calculating the effective potential
$V_{eff}$ by finding the entire $m^2$ spectrum, then taking their
square roots and expanding them in powers of $\mu$ and $v$ becomes
inefficient, since in the most general case this involves finding the
eigenvalues of mass matrices of very high dimension.

Instead, it is possible to make use of the sum over mass formula in
eqn(\ref{Vformula}) without explicitly diagonalizing the mass
matrix. Let us denote the  square of the mass matrix as $W=M^2$. Since
there is never any mixing between the  bosons, the fermions and the
ghosts, we can study their mass matrices separately.

In terms of $W$, the sum over mass formula becomes:

\begin{equation}\label{VF2}
V_{eff}^{1-loop}=-\frac{1}{2} tr ( \sqrt{W_b} - \sqrt{W_f} -
\sqrt{W_g} )
\end{equation}

The square root of $W$ can be defined unambigously by its expansion in
powers of  $\alpha /r^2$ in the Supergravity limit, as was discussed
in Section \ref{SimCase}.  Note that $M_b$ is defined to be the mass
matrix for {\it real} bosons. If it is taken  to be the mass matrix
for the complex bosons, then there will be an extra factor of  two in
front of $\sqrt{W_b}$.

\subsection{Simple Recipe for Mass Matrix}

In this subsection we will give a simple recipe for writing out $M^2$
for both the bosons and the fermions. The mass for the ghosts is
exactly the same as in  the simple case of Section \ref{SimCase}.

First of all, we should note that the mass of $\zeta^i$ and $\zeta^a$
are always  $\mu/3$ and $\mu/6$ respectively for $i=1,2,3$ and
$a=4,...,9$. The mass of all eight physical  degrees in $\eta$ is
always $\mu/4$. These are independent of the background $B$. Mixing
occurs only among the $z^I$ and among the $\theta$ and
$\overline{\theta}$. Hence in what  follows, we will denote the
component arising from say $\overline{z}^I z^J$ in the bosonic
Lagrangian simply as $(M^2)_{IJ}$ without mentioning $z$
explicitly. Note also that  $M^2$ is symmetric.

\subsubsection{Rules for Bosons}

1.  $(M^2)_{00}=r^2; \quad (M^2)_{ii}=r^2+\mu^2/3^2; \quad
(M^2)_{aa}=r^2+\mu^2/6^2;$\\ 2.  $\dot{x}^I=v_I$ mixes $z^0$ and $z^I$
$\Rightarrow \ (M^2)_{0I}=-2  v_I$\\ 3.  $x^1=b_1$ mixes $z^2$ and
$z^3$... etc $\Rightarrow \ (M^2)_{jk}=i\mu  \epsilon_{ijk} b_i$\\

Note that Rule 3 applies only to $z^i$ but not $z^a$. Such mixing is
the effect of  the Myers term in the Matrix Theory action.

\subsubsection{Rules for Fermions}

The mass matrix for the fermions can be written in a closed form:\\
\begin{equation}\label{Mfformula}
M^2=r^2+\mu^2/4^2+\sum_{I=1}^9 v_I \gamma_I + \sum_{i=1}^3\frac{i\mu
x^i}{4} \{\gamma_i, \gamma_{123}\}
\end{equation}

\section{The General Case}\label{GenCase}

Once the mass matrix squared $W=M^2$ is known, eqn(\ref{VF2}) can then
be used to  compute the 1-loop effective potential explicitly. In
accordance with our earlier  discussions, only terms up to order
$\alpha^3 \sim 16 \pi^5/M^9=\kappa_{11}^2$ are kept.  After restoring
all factors of $M^3R$, $N_p$, and $N_s$, the 0 and 1-loop effective
potential is given by:

\begin{multline} \label{VGauge2}
V_{eff}^{0,1-loop}=\frac{N_p}{2R} (\sum_{I=1}^9 v_I^2+ g_{++}) +
\frac{N_p N_s}{M^9  R^3} \Bigg \{ \frac{15(\sum_{I=1}^9 v_I^2)^2}{16
r^7}-\frac{\mu^2 \sum_{i=1}^3  v_i^2}{96  r^5} - \frac{7 \mu^2
\sum_{a=4}^9 v_a^2}{96  r^5}\\  +\frac{15\mu^2}{32  r^7}\left [
\sum_{i=1}^3 x_i^2\left(-\sum_{i=1}^3  v_i^2+\sum_{a=4}^9
v_a^2\right)+2 (\sum_{i=1}^3 x_i v_i)^2 \right] \Bigg \}\\
+\frac{\mu^4N_pN_s}{R^3M^9}\frac{1}{768r^7}\Bigg \{
32\left[\sum_{i=1}^3(x^i)^2\right]^2+\left[\sum_{a=4}^9(x^a)^2\right]^2-12\sum_{i=1}^3(x^i)^2\cdot\sum_{a=4}^9(x^a)^2\Bigg
\}
\end{multline}
This is the equation to be compared with the Supergravity result.
Notice the effective potential has manifest $SO(3)\times SO(6)$
symmetry, as should  be expected from the symmetry of the original
Matrix Theory action. Just as in flat space \cite{TRCB}, one should be
able to recast this 1-loop effective potential in the form $T^{\mu
\nu} G_{\mu \nu}$. A comparison with the Supergravity side will indeed
confirm this, as this  is precisely the form of the effective
potential on the Supergravity side as derived in Appendix B.

Having computed the effective potential on the Matrix Theory side, the
next step will be to compare it with the result from a Supergravity
calculation. Before this could be done, the issue of gauge choice has
to  be addressed.

It is necessary to make a gauge choice when solving the Einstein
equations. A gauge choice corresponds to a choice of the coordinate
system one  uses to describe the physics. On the Matrix Theory side,
such a choice of  coordinates was made right from the very beginning:
The action in  eqn(\ref{Maction}) was written in coordinates that made
the $SO(3)\times  SO(6)$ symmetry manifest. Before a comparison is
possible, a  corresponding choice of coordinates, i.e. a choice of
gauge has to be  made on the Supergravity side.

A comparison of the above equation with the general expression for
$V_{eff}$ in  eqn(\ref{VSug}) will in the end determine the correct
gauge choice for the  Supergravity computation. There will be a
further discussion about gauge choice in the  Supergravity section.

\section{The Supergravity Effective Potential}\label{sec:sugra}

To find the two-body effective
action, one only  needs to solve for the metric perturbation caused by
the source graviton at the linear order  ($\sim\kappa_{11}^2$).

The action is given by:
\begin{equation}
S=S_G+S_A+S_P
\end{equation}
$S_G$ is the Einstein action for the metric:
\begin{equation}
S_G=\frac{1}{\kappa_{11}^2}\int d^{11}x \sqrt{|g|}R
\end{equation}
$S_A$ is the action for the three-form:
\begin{equation}
S_A=-\frac{2}{\kappa_{11}^2}\int d^{11}x  \left\{
\frac{\sqrt{|g|}}{2\cdot 2\cdot 4!}F^{\mu\nu\lambda\xi}F_{\mu\nu\lambda\xi} +\frac{1}{12}
\frac{1}{3!(4!)^2}\epsilon^{\mu_1\ldots \mu_{11}}
A_{\mu_1\mu_2\mu_3}F_{\mu_4\ldots \mu_7}F_{\mu_8\ldots \mu_{11}}\right\}
\end{equation} 
$S_P$ is the action for the source graviton (the subscript $P$ means
"particle"):
\begin{equation}
S_P=C_P\frac{1}{2}\int_{-\infty}^{+\infty}d\xi
\left(\frac{1}{\beta(\xi)}g_{\mu\nu}(y) \frac{dy^\mu}{d\xi}
\frac{dy^\nu}{d\xi}-\beta(\xi)m^2 \right)
\end{equation}
with $C_P$ being some constant.

The above action gives the equations of motion for the metric, the
3-form field, and  the source graviton, all listed below.  \\ The
Einstein equation:
\begin{equation}
R_{\mu\nu}-\frac{1}{2}Rg_{\mu\nu}
=\kappa_{11}^2\left(\left[T_{\mu\nu}\right]_A+\left[T_{\mu\nu}\right]_P
\right)
\end{equation}
The Maxwell equation:
\begin{equation}
\partial_\mu\left(\sqrt{|g|}F^{\mu\nu\lambda\xi}\right)
-\frac{1}{1152}\epsilon^{\nu\lambda\xi\rho_1\ldots\rho_8}F_{\rho_1\ldots\rho_4}F_{\rho_5\ldots\rho_8}=0
\end{equation}
The geodesic equation:
\begin{equation}
\frac{d^2y^\mu}{d\xi^2}
+\Gamma^\mu_{\rho\nu}(y)\frac{dy^\rho}{d\xi}\frac{dy^\nu}{d\xi}=0
\end{equation}
$\left[T_{\mu\nu}\right]_A$ and $\left[T_{\mu\nu}\right]_P$ are the
stress tensors obtained by varying $S_A$ and $S_P$ w.r.t. the metric,
given below:
\begin{equation}
\left[T_{\mu\nu}\right]_A =\frac{1}{12\kappa_{11}^2}\left( F_{\mu
\lambda\xi\rho}F_\nu^{\lambda\xi\rho}-\frac{1}{8}g_{\mu\nu}F^{\rho\sigma\lambda\xi}F_{\rho\sigma\lambda\xi}\right)
\end{equation}
\begin{equation}
\left[T_{\mu\nu}\right]_P(x)
=\frac{C_P}{2}\frac{1}{\sqrt{|g(x)|}}g_{\mu\rho}(x)g_{\nu\lambda}(x)
\int_{-\infty}^{+\infty} d\xi
\frac{1}{\beta(\xi)}\frac{dy^\rho(\xi)}{d\xi}
\frac{dy^\lambda(\xi)}{d\xi}  \delta^{(11)}(x-y(\xi))
\end{equation}
 
Setting $C_P$ to zero means the absence of the source graviton. In
this case, a solution to the above equations of motion is the
pp-wave background.   The metric $g_{\mu\nu}$ and the 4-form field
strength are given by:

\begin{equation}
g_{+-}=1,\ g_{++}
=-\mu^2\left[\frac{1}{9}\sum_{i=1}^3(x^i)^2+\frac{1}{36}
\sum_{a=4}^9(x^a)^2 \right],\ g_{AB}=\delta_{AB}
\end{equation}

\begin{equation}
F_{123+}=\mu
\end{equation}
In our convention, $\mu,\nu,\rho,\ldots$ take the values
$+,-,1,\ldots, 9$;  $A,B,C,\ldots$ take the values $1,\ldots,9$;
$i,j,k,\ldots$ take the values  $1,\ldots,3$; and $a,b,c,\ldots$ take
the values $4,\ldots,9$

The introduction of a source graviton, i.e. a non-zero $C_P$, perturbs
the  above pp-wave solution:
\begin{equation}
g_{\mu\nu}\longrightarrow g_{\mu\nu}+h_{\mu\nu}\equiv G_{\mu\nu};\
F_{\mu\nu\rho\sigma}\longrightarrow
F_{\mu\nu\rho\sigma}+f_{\mu\nu\rho\sigma}\nonumber
\end{equation}
It suffices to solve the geodesic equation at the zeroth order of
$C_P$, which gives a  solution
\begin{equation}
x^+=\xi,\ x^-=0,\ x^A=0\nonumber
\end{equation} 
and the corresponding stress tensor of the source graviton is then:
\begin{equation}
\left[T_{\mu\nu}\right]_P(x)
=p^+g_{\mu+}g_{\nu+}\delta(x^-)\prod_{A=1}^9\delta(x^A)
\end{equation}
where $p^+\equiv \frac{C_P}{2\beta_0}$ is a constant (note that
$\beta(\xi)$ is a constant $\beta_0$ for a geodesic) and in what
follows  we will use $p^+$ instead of $C_P$. Note that the order of
$\kappa_{11}^2$ is the same  as the order of $p^+$.  Also note that
the only non-vanishing component of  $\left[T_{\mu\nu}\right]_P$ is
$\left[T_{--}\right]_P=p^+\delta(x^-)\prod_{A=1}^9\delta(x^A)$.

In what follows we will integrate everything over the $x^-$ direction,
thus  getting rid of $\delta(x^-)$ and derivatives w.r.t $x^-$.   On
the Matrix Theory side, the effective potential was only computed up
to 1-loop. In  Supergravity language, that means we are only looking
at order $\kappa_{11}^2$. To  find the effective potential on the
Supergravity side up to this order, we need only  the linearized
(i.e. to the linear order of $p^+$) Einstein equation and Maxwell
equation.

We consider static solutions which has no $x^+$ dependence. Also we
restrict our  attention to metric and gauge field perturbations that
go to zero at infinity.  The linearized Einstein equation in 11
dimension is:
\begin{equation}
\delta R_{\mu\nu} =\kappa_{11}^2\left[\delta T_{\mu\nu}
+\frac{1}{9}g_{\mu\nu}\left(T^{\alpha\beta}h_{\alpha\beta}-g^{\alpha\beta}
\delta T_{\alpha\beta} \right)  \right]\equiv{\cal T}_{\mu\nu}
\end{equation} 
where the perturbation to the total stress tensor is given by
\begin{equation}
\delta T_{\alpha\beta}=\left[\delta T_{\alpha\beta}\right]_A
+\left[T_{\alpha\beta}\right]_P
\end{equation}
$\left[\delta T_{\alpha\beta}\right]_A$ is the perturbation to the
stress tensor of the gauge field, which is to be expressed in terms
of the perturbation to the field strength.

First look at the $(--)$ component of the Einstein equation:

\BE \delta R_{--}=-\frac{1}{2}\sum_{A=1}^9
\frac{\partial^2h_{--}}{\partial x^A\partial x^A} \EE

and \BE {\cal T}_{--}=\kappa_{11}^2\delta T_{--}=\kappa_{11}^2 \left
[T_{--}\right]_P=\kappa_{11}^2p^+\prod_{A=1}^9\delta(x^A) \EE where
$\left[\delta T_{--}\right]_A=0$ (as can be readily verified) has been
used.

This gives \BE
h_{--}=\frac{\kappa_{11}^2p^+}{\pi^4}\frac{15}{16}\frac{1}{|\vec{x}|^7}\label{hmm}
\EE where we use $\vec{x}$ to denote the 9-dimensional vector in the
transverse  directions.

The $(-A)$ component of the Einstein equation is, \BE \delta
R_{-A}=-\frac{1}{2}\sum_{B=1}^9 \frac{\partial^2h_{-A}}{\partial
x^B\partial x^B}+\frac{1}{2}\sum_{B=1}^9
\frac{\partial^2h_{-B}}{\partial x^A\partial x^B} \EE and \BE {\cal
T}_{-A}=0 \EE which gives \BE h_{-A}=0\label{hma} \EE

Now we look at the linearized Maxwell equation, in terms of the gauge
potential  perturbation $a_{\mu\nu\rho}$ (note
$f_{\lambda\mu\nu\rho}=\partial_\lambda  a_{\mu\nu\rho}-\partial_\mu
a_{\nu\rho\lambda}+\partial_\nu  a_{\rho\lambda\mu}-\partial_\rho
a_{\lambda\mu\nu}$). We choose to work in the  ``Lorentz gauge'' where
$\sum_{D=1}^9 \partial_D a_{\mu\nu D}=0$.  The upper $(AB+)$ component
of the Maxwell equation gives: \BE \sum_{D=1}^9 \partial^2_D a_{AB-}-
\sum_{D=1}^9  \partial_D\left[h_{--}F_{DAB+}\right]=0 \EE Using the
expression for $h_{--}$ that we just found, we have:  \BE
a_{ij-}=\frac{\mu\kappa_{11}^2p^+}{\pi^4}\
\frac{15}{32}\sum_{k=1}^3\epsilon_{ijk}\frac{x^k}{\left|\vec{x}\right|^7}
\EE  while all other $a_{AB-}$'s vanish.  This gives the field
strength: \BE f_{-ijk}&=&\frac{\mu\kappa_{11}^2p^+}{\pi^4}\
\frac{15}{32}\epsilon_{ijk}\left[7\frac{\sum_{i=1}^3(x^i)^2}{\left|\vec{x}\right|^9}-3\frac{1}{\left|\vec{x}\right|^7}\right]\nonumber\\
f_{-ijb}&=&\frac{\mu\kappa_{11}^2p^+}{\pi^4}\
\frac{15}{32}\sum_{k=1}^3\epsilon_{ijk}\left[7\frac{x^kx^b}{\left|\vec{x}\right|^9}\right]\label{fm}
\EE

Next consider the upper $(ABC)$ component of the Maxwell
equation. Using the fact that  $h_{-A}=0$ and $a_{AB-}=0$ except for
$a_{ij-}$, we have: \BE \sum_{D=1}^9 \partial^2_D a_{ABC}=0 \EE
hence, all $a_{ABC}=0$.  Now the $(A+-)$ component. Using  $h_{-A}=0$
we get \BE \sum_{D=1}^9 \partial^2_D a_{A-+}=0 \EE  thus $a_{A-+}=0$.
Now we go back to look at the $(+A)$ component of the Einstein
equation. Using $h_{-A}=0$, we  get  \BE \delta
R_{+A}=-\frac{1}{2}\sum_{B=1}^9 \frac{\partial^2h_{+A}}{\partial
x^B\partial x^B}+\frac{1}{2}\sum_{B=1}^9
\frac{\partial^2h_{+B}}{\partial  x^A\partial x^B} \EE Using
$a_{A-+}=0$, $a_{ABC}=0$, and $h_{-A}=0$, we get \BE {\cal T}_{+A}=0
\EE So we conclude that \BE h_{+A}=0\label{hpa} \EE

Now consider the $(+-)$ component of the Einstein equation \BE \delta
R_{+-}=-\frac{1}{2}\sum_{A=1}^9 \frac{\partial^2h_{+-}}{\partial
x^A\partial x^A} +\frac{1}{2}\sum_{A=1}^9  \frac{\partial
g_{++}}{\partial x^A}\frac{\partial h_{--}}{\partial x^A}  \EE and \BE
{\cal T}_{+-}=\frac{1}{6}\left(\mu^2h_{--}-\mu f_{-123}\right) \EE

In writing ${\cal T}_{+-}$, we made use of the following equations:
\BE \left[\delta
T_{+-}\right]_A&=&\frac{\mu^2}{4\kappa_{11}^2}h_{--}\nonumber\\
\left[\delta
T_{ij}\right]_A&=&\frac{1}{4\kappa_{11}^2}\delta_{ij}\left(-2\mu
f_{-123}-\mu^2h_{--}\right)\nonumber\\ \left[\delta
T_{bc}\right]_A&=&\frac{1}{4\kappa_{11}^2}\delta_{bc}\left(2\mu
f_{-123}+\mu^2h_{--}\right)\nonumber\\ \left[\delta
T_{ib}\right]_A&=&-\
\frac{\mu}{4\kappa_{11}^2}\sum_{j,k=1}^3\epsilon_{ijk}f_{-jkb} \EE

Solving this Einstein equation we have: \BE
h_{+-}=-\frac{\mu^2\kappa_{11}^2p^+}{\pi^4}
\left[\frac{5}{64}\frac{\sum_{i=1}^3(x^i)^2}{|\vec{x}|^7}
+\frac{1}{192}\frac{1}{|\vec{x}|^5} \right]\label{hpm} \EE The $(AB)$
component of the Einstein equation reads: \BE \delta
R_{AB}&=&-\frac{1}{2} \left[\sum_{C=1}^9\frac{\partial^2h_{AB}}
{\partial x^C\partial x^C}-\sum_{C=1}^9\frac{\partial^2h_{AC}}
{\partial x^B\partial x^C}-\sum_{C=1}^9\frac{\partial^2h_{BC}}
{\partial x^A\partial x^C}+\sum_{C=1}^9\frac{\partial^2h_{CC}}
{\partial x^A\partial x^B}+2\frac{\partial^2h_{+-}} {\partial
x^A\partial x^B}  \right] \nonumber\\ &+&\frac{1}{4}
\left[2h_{--}\frac{\partial^2g_{++}} {\partial  x^A\partial
x^B}+2g_{++}\frac{\partial^2h_{--}} {\partial   x^A\partial
x^B}+\frac{\partial g_{++}}{\partial x^A}\frac{\partial
h_{--}}{\partial x^B}+\frac{\partial g_{++}}{\partial
x^B}\frac{\partial  h_{--}}{\partial x^A} \right]  \EE and \BE {\cal
T}_{ij}&=&-\ \frac{1}{3}\delta_{ij}\left(2\mu
f_{-123}+\mu^2h_{--}\right)\nonumber\\  {\cal
T}_{bc}&=&\frac{1}{6}\delta_{bc}\left(2\mu
f_{-123}+\mu^2h_{--}\right)\nonumber  \\ {\cal T}_{ib}&=&-\
\frac{\mu}{4}\sum_{j,k=1}^3\epsilon_{ijk}f_{-jkb} \EE

So far the need to make a gauge choice for the metric has not
arisen. Now to solve for  $h_{AB}$ we must make a gauge choice for the
metric.  Let $G^{\rho\sigma}$ and $\Gamma^\mu_{\rho\sigma}$ denote the
complete  inverse metric and Christoffel symbol respectively (by
"complete", we mean  they include both the unperturbed and perturbed
part). We shall fix the  gauge by specifying
$G^{\rho\sigma}\Gamma^\mu_{\rho\sigma}$.

As can be easily verified, \BE
G^{\rho\sigma}\Gamma^+_{\rho\sigma}&=&\sum_{C=1}^9\partial_Ch_{-C}=0\nonumber
\\
G^{\rho\sigma}\Gamma^-_{\rho\sigma}&=&\sum_{C=1}^9\left(-h_{-C}\partial_Cg_{++}+\partial_Ch_{+C}-g_{++}\partial_Ch_{-C}\right)=0\nonumber\\
G^{\rho\sigma}\Gamma^A_{\rho\sigma}&=&\sum_{C=1}^9\partial_Ch_{AC}-\frac{1}{2}\partial_A\left(\sum_{C=1}^9h_{CC}+2h_{+-}-g_{++}h_{--}\right)
\EE so we need to specify $G^{\rho\sigma}\Gamma^A_{\rho\sigma}$ to fix
the gauge.

Using the above expressions for $G^{\rho\sigma}\Gamma^A_{\rho\sigma}$,
we can rewrite  $\delta R_{AB}$ as \BE \delta R_{AB}=-\frac{1}{2}
\left[\sum_{C=1}^9\frac{\partial^2h_{AB}}{\partial  x^C\partial
x^C}-\frac{\partial \left(G^{\rho\sigma}\Gamma^A_{\rho\sigma}
\right)}{\partial x^B}-\frac{\partial
\left(G^{\rho\sigma}\Gamma^B_{\rho\sigma} \right)}{\partial
x^A}+\frac{1}{2}\left(\frac{\partial g_{++}}{\partial
x^A}\frac{\partial h_{--}}{\partial x^B}+\frac{\partial
g_{++}}{\partial    x^B}\frac{\partial h_{--}}{\partial x^A}  \right)
\right]   \EE

In general relativity we often use the ``harmonic gauge'' where we set
$G^{\rho\sigma}\Gamma^A_{\rho\sigma}=0$ (which is satisfied by the
unperturbed pp-wave background). Here, however, we shall opt for a
different gauge.

As derived in the Appendix B, the effective potential is given by:
\begin{eqnarray} \label{VSug}
V_{eff}= \frac{N_p}{R}\Bigg\{\frac{1}{2}\bigg
[v^2+g_{++}+h_{++}+g_{++}\left(\frac{1}{4}g_{++}h_{--}-h_{+-}\right)\nonumber\\
+\sum_A [2h_{+A}-h_{-A}(v^2+g_{++})]v^A+\sum_{A,B}h_{AB}v^Av^B \bigg ]\nonumber\\
+\frac{1}{8}h_{--}v^4
-\frac{1}{2}v^2\left(h_{+-}-\frac{1}{2}g_{++}h_{--}  \right)\Bigg\}
\end{eqnarray}
where $N_p$ is the number of D0-branes forming the probe graviton, and
$v^A\equiv\dot{x}^A$, $v^2\equiv\sum_{A=1}^9 (v^A)^2$. As
$h_{+A},h_{-A}$ all vanish,  they simply drop out of the effective
potential.

The computation on Matrix Theory side in section~\ref{GenCase} tells
us that in the effective potential there are no terms of the form
$v^av^b$ for $a\neq  b$, nor are there terms of the form
$v^iv^a$. This suggests we choose the gauge such  that $h_{ab}\propto
\delta_{ab}$, and $h_{ia}=0$.  To make $h_{ab}\propto \delta_{ab}$, we
set:
\begin{eqnarray}
G^{\rho\sigma}\Gamma^a_{\rho\sigma}
=\frac{1}{2}h_{--}\partial_ag_{++}\label{gaugechoice1}
\end{eqnarray}
then, to make $ h_{ia}=0$, we set: \BE
\partial_b\left(G^{\rho\sigma}\Gamma^i_{\rho\sigma}\right)=\frac{1}{2}\partial_ig_{++}\partial_bh_{--}-\frac{\mu}{2}\epsilon_{ijk}f_{-jkb}\nonumber
\EE which implies
\begin{eqnarray}
G^{\rho\sigma}\Gamma^i_{\rho\sigma}
=\frac{35}{96}\frac{\mu^2\kappa_{11}^2p^+}{\pi^4}\frac{x^i}{|\vec{x}|^7}
\label{gaugechoice2}
\end{eqnarray}
In this gauge, the Einstein equation gives:
\begin{equation}
h_{ab}=\delta_{ab}\frac{\mu^2\kappa_{11}^2p^+}{\pi^4} \frac{1}{96}
\left[\frac{15}{2}\frac{\sum_{k=1}^3(x^k)^2}{|\vec{x}|^7}
-\frac{1}{|\vec{x}|^5} \right]\label{hab}
\end{equation}
\begin{equation}
h_{ij}=\delta_{ij}\frac{\mu^2\kappa_{11}^2p^+}{\pi^4} \frac{1}{96}
\left[ -15 \frac{\sum_{k=1}^3(x^k)^2}{|\vec{x}|^7}
+\frac{1}{2}\frac{1}{|\vec{x}|^5} \right]
+\frac{\mu^2\kappa_{11}^2p^+}{\pi^4}\frac{15}{64}\frac{x^ix^j}{|\vec{x}|^7}\label{hij}
\end{equation} 

Now let us look at the upper $(AB-)$ component of the Maxwell
equation.  It gives the following equations:\BE &\
&\sum_{D=1}^9\partial_D^2a_{ij+}-g_{++}\sum_{D=1}^9\partial_D^2a_{ij-}-\sum_{D=1}^9\partial_Dg_{++}\left(\partial_Da_{ij-}+\partial_ia_{jD-}+\partial_ja_{Di-}\right)\nonumber\\
&\
&+\mu\sum_{k=1}^3\epsilon_{ijk}\Bigg\{-\sum_{D=1}^9\partial_Dh_{Dk}+\sum_{m=1}^3\left(\partial_mh_{mk}-\partial_kh_{mm}\right)+\partial_k\left[\frac{1}{2}\left(g_{++}h_{--}+\sum_{D=1}^9h_{DD}\right)\right]\Bigg\}\nonumber\\
&\ &=0 \EE \BE \sum_{D=1}^9\partial_D^2a_{bc+}=0 \\
\sum_{D=1}^9\partial_D^2a_{ib+}=0 \EE Solving them gives: \BE
a_{ij+}&=&\frac{\mu^3\kappa_{11}^2p^+}{\pi^4}\left(\sum_{k=1}^3\epsilon_{ijk}x^k\right)\frac{1}{384\left|\vec{x}\right|^7}\left[-29\sum_{m=1}^3(x^m)^2+\sum_{a=4}^9(x^a)^2\right]\\
a_{bc+}&=&0\\ a_{ib+}&=&0 \EE They give the field strength: \BE
f_{+ijk}&=&\frac{\mu^3\kappa_{11}^2p^+}{\pi^4}\epsilon_{ijk}\frac{1}{384\left|\vec{x}\right|^9}\left[-58\sum_{m=1}^3(x^m)^2-3\sum_{a=4}^9(x^a)^2+149\sum_{m=1}^3(x^m)^2\cdot\sum_{a=4}^9(x^a)^2
\right]\nonumber\\
f_{+ijb}&=&\frac{\mu^3\kappa_{11}^2p^+}{\pi^4}\left(\sum_{k=1}^3\epsilon_{ijk}x^k\right)\frac{5}{384}\
\frac{x^b}{\left|\vec{x}\right|^9}\left[-41\sum_{m=1}^3(x^m)^2+\sum_{a=4}^9(x^a)^2\right]\label{fp}
\EE As can be easily checked, all the $a_{\mu\nu\rho}$ we have found
indeed satisfy the  Lorentz gauge.

Finally, we consider the $(++)$ component of the Einstein equation:
\BE \delta R_{++}&=&-\
\frac{1}{2}\sum_{A=1}^9\partial_A^2h_{++}+\frac{1}{2}\sum_{A,B=1}^9\partial_Ag_{++}\partial_Bh_{AB}-\frac{1}{4}\sum_{A,B=1}^9\partial_Ag_{++}\partial_Ah_{BB}\nonumber\\
&\
&+\frac{1}{2}\sum_{A,B=1}^9h_{AB}\partial_A\partial_Bg_{++}+\frac{1}{2}\sum_{A=1}^9\partial_Ag_{++}\partial_Ah_{+-}+\frac{1}{4}\sum_{A=1}^9g_{++}\partial_Ag_{++}\partial_Ah_{--}\nonumber\\
&\ &-\ \frac{1}{4}\sum_{A=1}^9h_{--}\left(\partial_Ag_{++}\right)^2
\EE and \BE {\cal T}_{++}=-\
\frac{\mu}{2}\left(2f_{+123}+\mu\sum_{i=1}^3h_{ii}\right)+\frac{\mu}{6}g_{++}\left(2f_{-123}+\mu
h_{--}\right) \EE From this we find \BE
h_{++}=\frac{\mu^4\kappa_{11}^2p^+}{\pi^4}\
\frac{1}{6912\left|\vec{x}\right|^7}\left\{116\left[\sum_{i=1}^3(x^i)^2\right]^2+2\left[\sum_{a=4}^9(x^a)^2\right]^2-17\sum_{i=1}^3(x^i)^2\cdot\sum_{a=4}^9(x^a)^2\right\}\label{hpp}
\EE

To summarize, the nonzero components of the metric perturbation are :
$h_{--}$  [eqn(\ref{hmm})], $h_{+-}$ [eqn(\ref{hpm})], $h_{ab}$
[eqn(\ref{hab})], $h_{ij}$  [eqn(\ref{hij})], $h_{++}$
[eqn(\ref{hpp})]; and the nonzero components of the field  strength
perturbation are: $f_{-ijk},f_{-ijb}$ [eqn(\ref{fm})] and
$f_{+ijk},f_{+ijb}$  [eqn(\ref{fp})].

Substituting the expressions for the metric into our formula for
$V_{eff}$ in  eqn(\ref{VSug}), averaging  $h_{\mu\nu}$ over
$x^-$(i.e. dividing by $2\pi R$), and noting that
$\kappa_{11}^2=\frac{16\pi^5}{M^9}$, $p^+=\frac{N_s}{R}$, we find
\begin{eqnarray}
V_{eff}&=&\frac{N_p}{2R}(v^2+g_{++})+\frac{15}{16}\frac{N_pN_s}{M^9R^3}
\frac{v^4}{|\vec{x}|^7}\nonumber\\ &+&\frac{\mu^2N_pN_s}{R^3M^9}
\Bigg\{\left[-\frac{1}{96}
\frac{1}{|\vec{x}|^5}-\frac{15}{32}\frac{(x^1)^2+(x^2)^2+(x^3)^2}
{|\vec{x}|^7}\right]\sum_{i=1}^3(v^i)^2
+\frac{15}{16}\frac{\sum_{i,j=1}^3x^ix^jv^iv^j}{|\vec{x}|^7}
\nonumber\\ &\ &\ \ \ \ \ \ \ \ \ \ \ \ \ \   +\left[-\frac{7}{96}
\frac{1}{|\vec{x}|^5}+\frac{15}{32}\frac{(x^1)^2+(x^2)^2+(x^3)^2}
{|\vec{x}|^7}\right]\sum_{a=4}^9(v^a)^2  \Bigg\}\nonumber\\
&+&\frac{\mu^4N_pN_s}{R^3M^9}\frac{1}{768\left|\vec{x}\right|^7}\left\{32\left[\sum_{i=1}^3(x^i)^2\right]^2+\left[\sum_{a=4}^9(x^a)^2\right]^2-12\sum_{i=1}^3(x^i)^2\cdot\sum_{a=4}^9(x^a)^2\right\}
\end{eqnarray} 
Comparison of the above formula with eqn(\ref{VGauge2}) on the Matrix
Theory side  shows exact agreement.

We would like to emphasize the approximation involved once again. We
treated the source graviton as a perturbation to the exact pp-wave
background, and the calculation was performed only to first order in
$p^+$. However the solution that we found for these linearized equations is exact in
$\mu$.

\section{Discussion and Future Directions} \label{sec:discussion}

In this paper, the effective potentials of Matrix Theory and
Supergravity describing  graviton scattering are computed, and they
are found to agree exactly at order  $\kappa_{11}^2$, up to quantum
corrections at short distances.  This provides evidence that the
Matrix Theory action proposed in \cite{BMN} describes  M-theory in the
pp-wave background. Furthermore, it implies the existence of a
nonrenormalization theorem protecting the terms studied against higher
loop  corrections on the Matrix Theory side. Such nonrenormalization
theorems have been  studied in flat space \cite{PSS1, PSS2, PSS3, HKS,
LOWE, NP, KM}. In these papers, the  $SO(9)$ symmetry in the
transverse space appears to be a crucial ingredient. In the  pp-wave
background, however, this is broken to $SO(3) \times SO(6)$, so it
will be  interesting to see how the result could be extended in this
case.

Our result at order $\mu^2$ agrees with Taylor and Van Raamsdonk's
proposal in \cite{TRCB} for  Matrix Theory in a weakly curved
background up to linear terms.  As mentioned in their discussion,
their proposal  is proven only in the case where the background is
produced by well-defined Matrix  Theory configurations. It is not the
case for the pp-wave background, so their  proposal for the Matrix
Theory in this background, while convincing, is not a proven
fact. Thus the result at $\mu^2$, i.e. terms linear in the background,
can be treated  as additional evidence for their proposal, similar to
the explicit calculation in \cite{TR}, this time with a nontrivial
$g_{++}$ metric component.

The result at order $\mu^4$ is beyond linear order in the background,
and hence is a  completely new result. In fact, our calculation
explicitly shows that there are no higher  powers of $\mu$ in the effective
action of the Supergravity side at the order  $\kappa_{11}^2$. On the
Matrix Theory side, higher powers of $\mu$ are also not  expected for
long distances. When they do appear, they are always accompanied by
higher powers  of  $\alpha/r^2$, which indicates that they are
corrections to Supergravity at short  distances.  However, as it
stands, the corrections for velocity dependent terms are  unreliable
because they are computed using the sum over mass formula, which is
exact  only for terms independent of velocity or terms proportional to
$\kappa_{11}^2$,  as is shown in the Appendix A. Evaluating these
corrections exactly requires going  beyond the sum over mass formula,
and an efficient way of handling the mass matrix  will be of use.

An obvious future direction is to push our Matrix Theory computation
to 2-loop. This can test whether a nonrenormalization
theorem exists for the terms  compared. A 2-loop computation is also
necessary if we are to extend our result to  three-body interactions.

In this paper, a very special bosonic background is chosen, with the
source graviton  sitting at the origin. It may be of interests to
generalise to the  case where neither of the gravitons are fixed at
the origin.


A very interesting prospect is to calculate scattering from a
transverse fivebrane,  recently constructed in \cite{MALD}. However,
the effect of a transverse fivebrane in this background may only be
visible through a higher loop computation on the Matrix Theory side.

\section*{Acknowledgments}
The authors would like to thank J. Schwarz for his advice and comments
at various stages of this project, and to Y. Okawa for his numerous
helpful comments and insights. They would also like to thank
T. McLoughlin and I. Swanson for many useful discussions. H. K. Lee
was supported by the Croucher Foundation in the duration of this
project.

\section*{Appendix A}\label{sec:appendixa}

\subsection*{Sum Over Mass Formula}

In this appendix we will prove the sum over mass formula in 
eqn(\ref{Vformula}). 

The effective action $\Gamma$ of a theory expanded upon a background $B$ 
in the background field method is given by:

\begin{equation}
\Gamma=\frac{1}{2} [Tr _{boson}\ ln(-\partial_{\tau}^2+W_b(\tau)) - Tr 
_{fermion}\ ln(-\partial_{\tau}^2+W_f(\tau)) - Tr 
_{ghost}\ ln(-\partial_{\tau}^2+W_{gh}(\tau))]
\end{equation}

Here $W=M^2$ is the mass matrix squared for the fluctuating fields, and 
the trace $Tr$ is over both the functional space and the field component 
indices (which are, besided the $U(2)$ indices, the spacetime indices 
$0,1,..,9$ for the bosons, and the 16 Dirac spinor indices for the 
fermions). 

Take the trace of the boson, for example:

\begin{eqnarray}
\Gamma_{boson}&=&\frac{1}{2} Tr _{boson}\ ln(-\partial_{\tau}^2+W(\tau)) 
\nonumber \\
&=&-\frac{1}{2} Tr\  \int_0^\infty \frac{ds}{s} 
exp[-s(-\partial_{\tau}^2+W(\tau))]
\end{eqnarray}

The trace over functional space can be computed using the ``plane-wave'' 
basis wavefunctions $|\omega>=\frac{1}{\sqrt{2\pi}}e^{-i\omega \tau}$, 
\begin{eqnarray}
\Gamma_{boson}=-\frac{1}{2} tr\ \int d\tau \int \frac{d\omega}{2\pi} 
\int_0^\infty 
\frac{ds}{s} e^{i\omega\tau} exp[-s(-\partial_{\tau}^2+W(\tau))] 
e^{-i\omega\tau}
\end{eqnarray}
The trace $tr$ is now over only the field component indices. If we define 
$V_{eff}$ by:
\begin{equation}
\Gamma=-\int d\tau V_{eff}
\end{equation}
Then, the bosonic part of $V_{eff}$ becomes:
\begin{equation}
V_{eff}(boson) = \frac{1}{2} tr\ \int \frac{d\omega}{2\pi} \int_0^\infty 
\frac{ds}{s} 
e^{i\omega\tau} exp[-s(-\partial_{\tau}^2+W(\tau))] e^{-i\omega\tau} \\
\end{equation}
The operator in the middle can be rewritten in the following way:
\begin{equation}
exp[-s(-\partial_{\tau}^2+W(\tau))]=X e^{-s W(\tau)} 
e^{+s\partial_{\tau}^2}
\end{equation}
Where $X$ is defined as:

\begin{eqnarray}
X&\equiv&exp[-s(-\partial_{\tau}^2+W(\tau))] e^{-s\partial_{\tau}^2} e^{+s 
W(\tau)}\nonumber\\
&=&1+ commutator\ terms \label{XDef}
\end{eqnarray}

The commutator terms give corrections to Supergravity, so for the purpose of this paper, which is to see whether Matrix model can reproduce Supergravity results, we can ignore them. This 
claim will be proven shortly, after the result from approximating $X=1$ is 
examined. In this approximation, we have:
\begin{eqnarray}
V_{eff}(boson)&=&\frac{1}{2} tr\ \int \frac{d\omega}{2\pi} \int_0^\infty 
\frac{ds}{s} 
exp[-s(\omega^2+W(\tau))] \nonumber \\
\Rightarrow \quad V_{eff}&=&-\frac{1}{2}tr \ M
\end{eqnarray}
Note that $M$, the square root of $W$, can be defined through its 
expansion in powers of 
$1/r$. 
Putting everything together, and minding the minus signs for the fermions 
and the ghosts, we 
get the sum over mass formula in eqn(\ref{Vformula}).
Now, we return to the claim made above, that the commutator terms in $X$ 
will not contribute to terms in the Supergravity limit.
To show this, we first write $X$ in a most general form:
\begin{equation}
X=\sum_{n,m}K[^m_n](W) s^n \partial^m
\end{equation}
Here $K[^m_n](W)$ is a general function of $W$ and its derivatives, and is {\it 
defined} by the above equation. Looking back at the definition of $X$ in 
eqn(\ref{XDef}), we see that $n$ counts the number of terms involved in forming the 
commutator, and $m$ is the number of derivatives not acting on $W$. For example, when 
$n=0$, it implies $m=0$, and $K[^0_0]=1$, corresponding to the approximation we made 
above. All the other values of $n$ correspond to commutator terms in $X$, and in 
particular, $K=0$ when $n=1$, because a commutator takes at least two terms.

Putting $X$ in terms of $K$ into $V_{eff}$, we will encounter the 
following factor inside the integrand:

\begin{eqnarray}
e^{i\omega \tau} \partial^m e^{-i\omega \tau} =\sum_{l=0}^m (^m_l) (-i \omega)^l \partial^{m-l}
=\sum_{l=0}^{[m/2]} (^m_{2l}) (- \omega^2)^l \partial^{m-2l}
\end{eqnarray}
$[m/2]$ is the biggest integer no larger than $m/2$. In the last line, we 
made use of the fact that $\omega$ will be integrated from $-\infty$ to 
$+\infty$ so that any odd functions in the integrand will give zero. As a 
result, only terms with even powers of $\omega$ are kept. Therefore, the 
effective potential becomes:

\begin{eqnarray}
V_{eff}(boson)&=&\frac{1}{4 \pi} \sum_{n,m} tr\ \int d\omega \int_0^\infty 
\frac{ds}{s} K[^m_n] s^n e^{i\omega \tau} \partial^m e^{-i\omega \tau} 
e^{-sW} e^{-s \omega^2} \nonumber \\
&=& \frac{1}{4 \pi} \sum_{n,m} \sum_{l=0}^{[m/2]} tr\ \int d\omega 
\int_0^\infty \frac{ds}{s} (^m_{2l}) (- \omega^2)^l e^{-s \omega^2} K[^m_n] s^n 
\partial^{m-2l} e^{-sW} \nonumber \\
&=& \frac{1}{4 \pi} \sum_{n,m} \sum_{l=0}^{[m/2]} tr\  \int_0^\infty 
\frac{ds}{s} (^m_{2l}) (\frac{\partial^l}{\partial s^l} 
\sqrt{\frac{\pi}{s}}) s^n K[^m_n] \partial^{m-2l} e^{-sW} \nonumber \\
&=& tr\ \sum_{n,m} \sum_{l=0}^{[m/2]} \frac{1}{4} (^m_{2l}) 
\frac{1}{\Gamma(1/2-l)} K[^m_n] \partial^{m-2l} \int_0^\infty \frac{ds}{s} 
s^{n-l-1/2} e^{-sW} \nonumber \\
&=& tr\ \sum_{n,m} \sum_{l=0}^{[m/2]} \frac{1}{4} (^m_{2l}) \frac{\Gamma 
(n-l-1/2)}{\Gamma (1/2-l)} K[^m_n] \partial^{m-2l} \frac{1}{W^{n-l-1/2}} 
\end{eqnarray}

Thus, the effective potential can be recasted in the following form:

\begin{equation}
V_{eff}=tr\ \sum_{n,m} \sum_{l=0}^{[m/2]} \frac{1}{4} (^m_{2l}) 
\frac{\Gamma (n-l-1/2)}{\Gamma (1/2-l)} \alpha^{2(n-l)-1} K[^m_n](W) 
\partial^{m-2l} \frac{1}{(\alpha^2 W)^{n-l-1/2}} 
\end{equation}
As before, $\alpha=1/(M^3 R)$. The reason these factors of $\alpha$ are 
inserted will be clear shortly. 

In a comparison of 1-loop Matrix Theory with Supergravity, the
relevant  terms on the Supergravity side are proportional to
$\kappa_{11}^2$, which  is of order $\alpha^3$ on the Matrix Theory
side. This means any terms of  higher powers of $\alpha$ are
irrelevant for such a comparison as they  represent only Matrix Theory
corrections to Supergravity, and finding them is not the  purpose of
this paper. In other words, to examine whether the Matrix  Theory can
reproduce Supergravity in the appropriate limit,  only terms up  to
$\alpha^3$ need to be kept.

It makes sense, therefore, to examine each factor in $V_{eff}$ and
count  the powers of $\alpha$ it contains. We begin with the mass
matrix squared. By inspection of the explicit expressions given in
setion \ref{MMC}, one sees that $W$ can always be written in
the following form:
\begin{eqnarray} \label{WN}
W &\sim& \frac{1}{\alpha ^2} r^2 + \frac{1}{\alpha} N \nonumber \\
\Rightarrow \alpha ^2 W &\sim& r^2 + \alpha N \nonumber \\
\Rightarrow (\alpha ^2 W)^k &\sim& 1 + \alpha N + (\alpha N)^2 + \cdots
\end{eqnarray}
For example, using eqn(\ref{Mfformula}), we have:
\begin{equation}
\alpha N_f=\alpha (\sum_{I=1}^9 v_I \gamma_I + \sum_{i=1}^3\frac{i\mu 
x^i}{4} \{\gamma_i, \gamma_{123}\})+\alpha^2 \mu^2/4^2 
\end{equation}
Similarly, $\alpha N_b$ can be constructed using the rules given in section \ref{MMC}, 
while $\alpha N_{gh}=0$. 
From the explicit expressions of $N$, it can be shown easily that $tr\  \alpha N=0$, 
$tr\ [(\alpha N)^2] =0 + O[\alpha^4]$, and $tr\ [(\alpha \partial N)^2] =0 + 
O[\alpha^4]$. These facts are related to the large number of supersymmetries of our system and will be of use shortly.
The last line in eqn(\ref{WN}) is a symbolic statement that for any $k$, whether 
positive or negative, $(\alpha ^2 W)^k$ will only give non-negative powers of 
$\alpha$. Another important point to note is that every $\alpha$ arising 
from $(\alpha ^2 W)^k$ is accompanied by a factor of $N$.
Now look at $K[^m_n]$:
Let $K[^m_n] = \sum_{p,q} K[^{m \ p}_{n \ q}]$
where $p$ is the number $\tau$-derivatives acting on $W$ inside $K$, and 
$q$ is the number of $W$ inside $K$.
By definition, we have:
\begin{equation}\label{Keq}
\frac{p+m}{2} + q = n
\end{equation}
For $n=0 \quad \Rightarrow K=1$;\\
For $n=1 \quad \Rightarrow K= 0$;\\
For $n\geq 2 \quad \Rightarrow K$ consists of commutators. In this case, 
we have:

\begin{equation}
\left \{ \begin{tabular}{l} $q<n$ \\ $\frac{p+m}{2}<n$ \end{tabular} \right. 
\end{equation}
For fixed $q$ and $n\geq 2$, $m$ and $p$ have the following extremal values:\\
$m_{min}=0 \quad \Rightarrow p_{max}=2(n-q)$\\
$p_{min}=1 \quad \Rightarrow m_{max}=2(n-q)-1$\\
The reason $p_{min}=1$ is that $\partial_\tau^2$ must act at least once on $W$ to give 
non-vanishing commutators like $[W,[W,[W, \dot{W}]]]$ in $K$.
Look at:
\begin{eqnarray}
\alpha^{2(n-l)-1} K[^m_n](W) \partial^{m-2l} \frac{1}{(\alpha^2 W)^{n-l-1/2}} 
&=&\sum_{p,q} \alpha^{2(n-l)-1} K[^{m \ p}_{n \ q}](W) \partial^{m-2l} \frac{1}{(\alpha^2 
W)^{n-l-1/2}} \nonumber \\
&=&\sum_{p,q} \alpha^{2(n-q) -1 -2l} K[^{m \ p}_{n \ q}](\alpha^2 W) \partial^{m-2l} 
\frac{1}{(\alpha^2 W)^{n-l-1/2}} \nonumber \\
&=&\sum_{p,q} \alpha^a K[^{m \ p}_{n \ q}](\alpha^2 W) \partial^{m-2l} \frac{1}{(\alpha^2 
W)^{n-l-1/2}}
\end{eqnarray}
where $a=2(n-q) -1 -2l$.
Noting $l \leq [m/2] \leq [m_{max}/2] = [n-q-1/2] = n-q-1$, we must have:
\begin{eqnarray}
l_{max}=n-q-1 \nonumber \\
\Rightarrow a\geq 1
\end{eqnarray}
From this derivation of the lower bound of $a$, we see that the equality holds only 
when $m=m_{max} = 2(n-q)-1$ and $l=l_{max}=n-q-1$. Then, eqn(\ref{Keq}) gives:
\begin{eqnarray}
m-2l=1\nonumber \\
p=p_{min}=1
\end{eqnarray}
A comparison with Supergravity with one-loop Matrix Theory means keeping terms only up 
to $\alpha^3 \sim \kappa_{11}^2$. Therefore, we need only consider the range of $a$ to 
be:
\begin{equation}
1 \leq a \leq 3
\end{equation}
For $a=3$:
There can be no factors of $\alpha N$ from $\alpha^2 W$ because they increase the 
powers of $\alpha$ beyond 3, hence taking us beyond the limit of Supergravity.
Without any factors of $\alpha N$, the effective potential is simply:
\begin{eqnarray}
V_{eff}&=&f(r) tr\ 1 \nonumber\\
\Rightarrow V_{eff} &\sim& tr\ 1 =0
\end{eqnarray}
Here, $tr$ is again the trace over the boson minus the trace over the fermions and the 
ghosts.
For $a=2$:
There can be at most one $\alpha N$, either from $K(\alpha^2 W)$ or $\partial^{m-2l} 
\frac{1}{(\alpha^2 W)^{n-l-1/2}}$ 
But for only one $N$, we have:
\begin{eqnarray}
V_{eff} &\sim& tr\ \partial^k N \nonumber\\
&=& \partial^k tr\  N \nonumber\\
&=& 0
\end{eqnarray}
For $a=1$:
Now it is possible to have $(\alpha N)^2$ coming from one of the following three 
cases:
(i) Both $(\alpha N)^2$ come from $\partial^{m-2l} \frac{1}{(\alpha^2 W)^{n-l-1/2}}$:
\begin{eqnarray}
V_{eff} &\sim& \alpha tr\ \partial^{m-2l}\left(\alpha N\right)^2 \nonumber \\
&\sim&\alpha \partial^{m-2l} tr\ \left(\alpha N\right)^2 \nonumber \\
&\sim& \alpha O[\alpha^4]\nonumber\\
&\sim& O[\alpha^5]
\end{eqnarray}
(ii)Both $(\alpha N)^2$ come from $K(\alpha^2 W)$:
Since we showed that $p=1$ when $a=1$, there is only one $\tau$-derivative acting on 
$W$ in $K$, we must have either:
(a) $V_{eff} \sim tr\ N^2$
or \ (b) $V_{eff} \sim tr\ ( N \partial_\tau N) \sim \frac{1}{2} \partial_\tau tr\ 
(N^2)$.
In either case, $V_{eff}=0+O[\alpha^5]$.\\
(iii) One $(\alpha N)$ comes from $K(\alpha^2 W)$ and one from $\partial^{m-2l} 
\frac{1}{(\alpha^2 W)^{n-l-1/2}}$:
We already showed that $m-2l=1$ when $a=1$, so we have:
\begin{equation}
V_{eff} \sim K \partial \frac{1}{(\alpha^2 W)^{n-l-1/2}}
\end{equation}
This implies either:
(a) $V_{eff} \sim tr\ (N \partial N)$
or \ (b) $V_{eff} \sim tr\ (\partial N \partial N)$\\
(a) is identical to case (iib) above. (b) is of order $\alpha^5$ using the fact mentioned before.  
This exhausts all cases contributing to terms up to order $\alpha^3 \sim 
\kappa_{11}^2$ in $V_{eff}$. In particular, we have shown that none of the commutator 
terms in $X$, corresponding to $K[^m_n]$ with  $n \geq 2$, contributes to terms relevant to Supergravity. This completes the proof of the claim made under equation (\ref{XDef}).

\section*{Appendix B}

\subsection*{Deriving $V_{eff}$}

The Lagrangian of the probe graviton moving in a curved spacetime with 
metric $G_{\mu\nu}=g_{\mu\nu}+h_{\mu\nu}$ is given by\footnote{This approach is borrowed from \cite{BBPT}.}:
\BE
{\cal L}=-m\sqrt{-G_{\mu\nu}\dot{x}^\mu\dot{x}^\nu} 
=-m\sqrt{-(g_{\mu\nu}+h_{\mu\nu})\dot{x}^\mu\dot{x}^\nu}
\EE
We make a Legendre transformation:
\BE
{\cal L'}={\cal L}-P^+_p\dot{x}^-\label{legendre}
\EE
where 
\BE
P^+_p=\frac{\delta{\cal L}}{\delta\dot{x}^-} 
=m\frac{1+h_{-\nu}\dot{x}^\nu}{\sqrt{-G_{\mu\nu}\dot{x}^\mu\dot{x}^\nu}} 
\EE
When we let $m\rightarrow 0$, this gives
\BE
G_{\mu\nu}\dot{x}^\mu\dot{x}^\nu =2\dot{x}^-+g_{++}+v^2+h_{\mu\nu} 
\dot{x}^\mu\dot{x}^\nu=0 
\EE
This is a quadratic equation for $\dot{x}^-$, which we will solve for 
$\dot{x}^-$, keeping only terms of lowest order of $p^+$. Recalling that all $h$ are at least of order $p^+$, we have:

\BE
\dot{x}^-=- \Bigg\{ \frac{1}{2}\bigg
[v^2+g_{++}+h_{++}+g_{++}\left(\frac{1}{4}g_{++}h_{--}-h_{+-}\right)\nonumber\\
+\sum_A [2h_{+A}-h_{-A}(v^2+g_{++})]v^A+\sum_{A,B}h_{AB}v^Av^B \bigg ]\nonumber\\
+\frac{1}{8}h_{--}v^4
-\frac{1}{2}v^2\left(h_{+-}-\frac{1}{2}g_{++}h_{--}  \right)\Bigg\}
\EE

Taking the limit $m\rightarrow 0$, the ${\cal L'}$ in eqn(\ref{legendre}) is simply $-P^+_p\dot{x}^-$, which is the effective potential $V_{eff}$ that we need. It contains the interaction between the probe and the source up to terms linear in $h_{\mu\nu}$. An alternative way of writing such interaction is:
\BE 
\frac{\delta {\cal L}}{\delta G_{\mu \nu}} h_{\mu \nu}= T^{\mu
\nu} h_{\mu \nu} 
\EE
This structure was used by Taylor and Van Raamsdonk \cite{TRCB} to identify the effective potentials on both sides.

\end{document}